\documentclass{main}
\usepackage{graphicx}      % include this line if your document contains figures
\usepackage{natbib}        % required for bibliography
\usepackage{amsmath,amssymb}
\begin{document}
\begin{frontmatter}

\title{Impact of Attitude and Bounded Rationality on Collective Behavioral Transitions} 

\thanks{This work was supported in part by the National Research Foundation, Singapore through its Medium Sized Center for Advanced Robotics Technology Innovation (CARTIN) under Project WP2.7, and in part by the project ``Humanizing the Sustainable Smart City (HiSS)'' within the KTH Digital Futures Research Program.}

\author[First,Third]{Chen Song} 
\author[Second]{Vladimir Cvetkovic} 
\author[Third]{Angela Fontan}
\author[First]{Rong Su}
\author[Third]{Karl H. Johansson}

\address[First]{School of Electrical and Electronic Engineering, Nanyang Technological University, Singapore, (e-mail: song0249@e.ntu.edu.sg; rsu@ntu.edu.sg)}
\address[Second]{Division of Resources, Energy, and Infrastructure, KTH Royal Institute of Technology, Stockholm, Sweden, (e-mail: vdc@kth.se)}
\address[Third]{Division of Decision and Control Systems, KTH Royal Institute of Technology, Stockholm, Sweden, (e-mail: \{angfon, kallej\}@kth.se)}

\begin{abstract}                % Abstract of 50--100 words
The theory of planned behavior (TPB) is one of the most influential frameworks in social psychology, stating that a person's behavior is driven by intention, which is primarily shaped by attitude, subjective norms, and perceived behavioral control. Despite its strong empirical support, TPB remains a static conceptual framework without explicit mathematical formulations that capture the temporal evolution of its components. To address this gap, we develop a dynamic agent-based modeling framework that integrates the core principles of TPB with a behavior-to-attitude feedback mechanism. Specifically, we define behaviors based on their feedback effects on attitude and examine when the population undergoes collective transitions by either adopting a beneficial behavior or rejecting a harmful one. Results from our model demonstrate that collective transitions can be effectively controlled by adjusting two key behavioral parameters that reflect agents' attitude influence and decision rationality. These findings provide quantitative insights on TPB, highlighting the key factors that drive collective behavioral transitions and the need for further socio-psychological case studies.   
\end{abstract}


\end{frontmatter}
%===============================================================================

\section{Introduction} \label{sec1}
\begin{figure}
\begin{center}
\includegraphics[width=8.4cm]{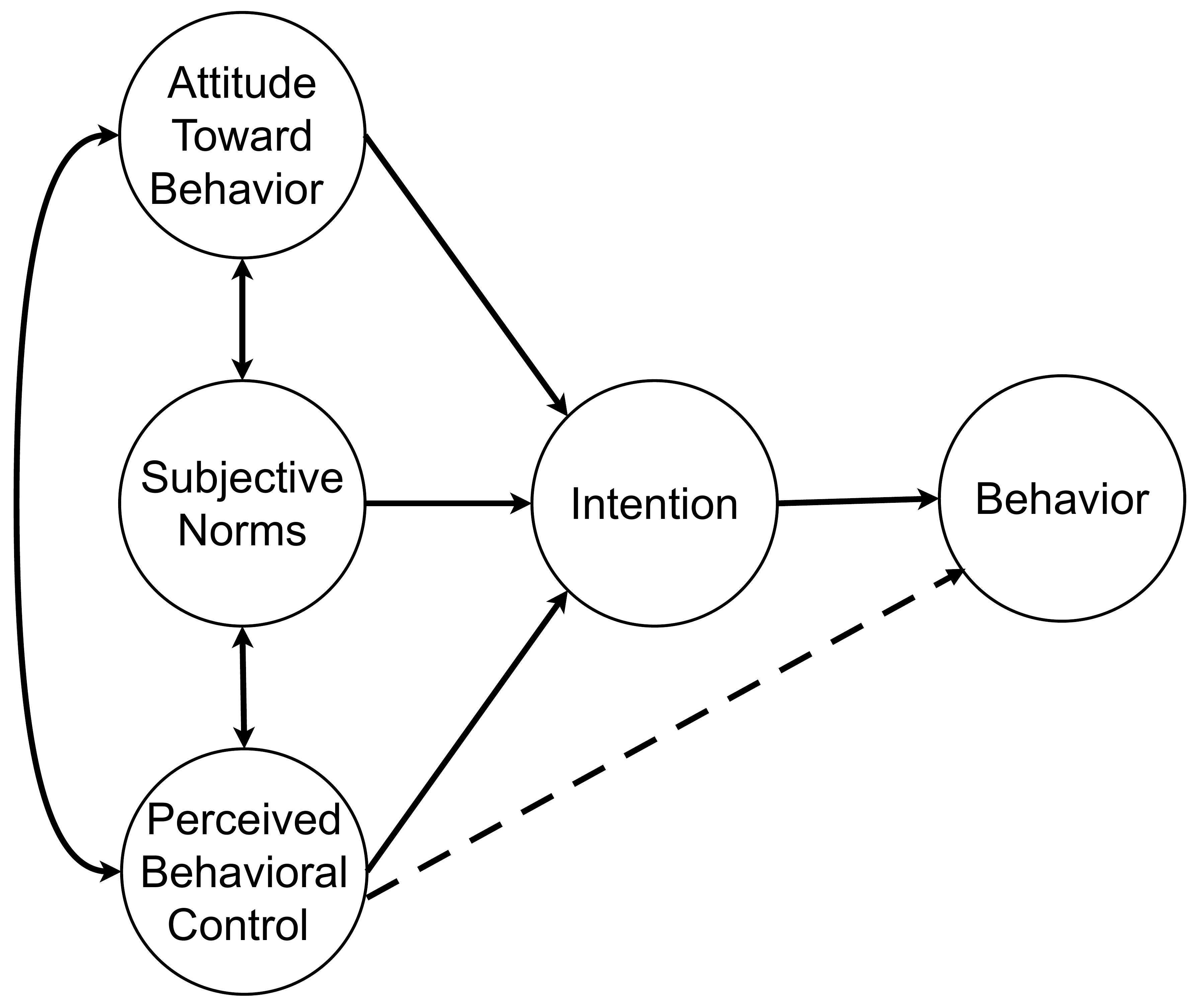}    % The printed column width is 8.4 cm.
\caption{Theory of planned behavior \citep{TPB}.} 
\label{fig0}
\end{center}
\end{figure}

Human behavior has been extensively studied in diverse disciplines, including behavioral economics, cognitive science, and social psychology. Among the various literature, the theory of planned behavior (TPB) \citep{TPB} serves as one of the most influential and widely adopted theoretical frameworks. As shown in Fig.~\ref{fig0}, TPB captures the role of personal attitude, subjective norms, and perceived behavioral control (i.e., cognitive limitations) in shaping one's intention and behavior. According to TPB, individuals are more likely to perform a behavior if they evaluate it positively, perceive social support, and feel capable of implementing it. A comprehensive review of TPB applications within the sustainability domain is provided in \cite{TPB_application}. Despite its strong empirical foundation, TPB remains conceptual and there has been a lack of dynamic, quantitative framework reflecting its core principles. In addition, TPB assumes a unidirectional influence from attitude to behavior without accounting for the fact that one's behavior could, in turn, affect their attitude \citep{cognitive_dissonace}. To address these limitations, we develop a dynamic agent-based modeling framework, which is grounded in TPB and incorporates a behavior-to-attitude feedback mechanism inspired by the psychophysical numbing principle \citep{psycho_numb_1}.       

Building on this framework, we examine how two key behavioral parameters, which reflect the influence weight of personal attitude on one's intention and the level of bounded rationality in decision-making, shape agents' collective behavioral transitions. In particular, we consider two types of behavior distinguished by their feedback effects on attitude: one with positive consequences that increase agents' attitude (beneficial type), and the other with negative consequences that decrease it (harmful type). Under different initial settings, we observe when the population transitions to adopt the beneficial behavior or reject the harmful one, and analyze how the two parameters govern the extent and speed of these transitions.      

The results indicate that the attitude influence weight plays a critical role in shaping collective transitions: a larger influence weight strengthens the reinforcement of beneficial behaviors and reduces conformity to subjective norms of harmful behaviors, facilitating transitions in both cases. In contrast, the level of bounded rationality exhibits a two-sided effect: a lower rationality level (i.e., increased randomness) in decision-making promotes transitions when initial conditions are unfavorable to them, but hinders them when the population is already inclined toward the transitions. These findings highlight the importance of personal attitude and the neutral role that bounded rationality plays in the social transition process, consistent with established socio-psychological evidence.

The rest of this paper is organized as follows. Section~\ref{sec2} provides a complete description of our agent-based model. Section~\ref{sec3} presents the main results followed by a detailed discussion. Section~\ref{sec4} draws conclusions and suggests directions for future work.

\section{Model Description} \label{sec2}
In this section, we provide a detailed description of the agent-based model. We first define the state variables for each agent and explain their interpretations. Then, we outline the model's sequential updating mechanism and specify the updating rule for each state variable along with their underlying design concepts. 

\subsection{Definition of State Variables} \label{sec2.1}
Consider a group of \(n\) agents, indexed by \(V=\{1,\dots,n\}\). Each agent \(i \in V\) is characterized by four state variables: namely attitude \(x_i\), intention \(z_i\), behavioral probability \(p_i\), and action \(y_i\), where \(x_i, z_i,\) and \(p_i\) are continuous variables defined on the interval \([0,1]\), and \(y_i\) is a binary variable taking values of either \(0\) or \(1\). The model is discretized in time, with all state variables updated at each time step \(t \in \{0,1,2,\dots\}\).  

The attitude \(x_i \in [0,1]\) reflects agent \(i\)'s internal evaluation of a certain behavior, consistent with the definition in the TPB \citep{TPB}. A large \(x_i\) indicates that agent \(i\) perceives the behavior as pleasant or satisfactory, whereas a small \(x_i\) implies that agent \(i\) holds a negative evaluation of it. This notion closely resembles the concept of ``value'' defined in the Prospect Theory (PT) \citep{PT}, where decision-makers assign numerical value to outcomes based on their subjective evaluation. 
%The attitude updating rule is also inspired by the value function proposed in \cite{PT}, to be explained in Sec~\ref{sec2.2}.  

The intention \(z_i \in [0,1]\) indicates agent \(i\)'s strength of motivation to perform the behavior. A large \(z_i\) implies a strong willingness of agent \(i\) to perform the behavior, while a small \(z_i\) indicates the contrary. According to the TPB, one's intention is mainly determined by its attitude and subjective norms, to be discussed in more detail later.  

%Since \(z_i\) is defined on a normalized scale of \([0,1]\) and each agent either performs or rejects the behavior, the value \(1-z_i\) can be naturally interpreted as the agent's intention to reject the behavior. Accordingly, when \(z_i>1-z_i\), i.e., \(z_i>0.5\), the agent is more likely to perform the behavior than to reject it.   

The behavioral probability \(p_i \in [0,1]\) denotes the likelihood that agent \(i\) performs the behavior. It has been well established in social psychology that a stronger intention leads to a higher probability of implementing that behavior \citep{Ajzen_Fishbean}. Thus, \(p_i\) is formulated as a monotonically increasing function of \(z_i\) in our model, such that \(p_i>0.5\) when \(z_i>0.5\) and \(p_i<0.5\) when \(z_i<0.5\).

Finally, the action \(y_i \in \{0,1\}\) represents the agent's realized behavior, where \(y_i=1\) or \(0\) indicates that agent \(i\) actually performs or rejects the behavior, with probability \(\mathbb{P}(y_i=1)=p_i\) and \(\mathbb{P}(y_i=0)=1-p_i\), respectively. In other words, \(y_i\) is a Bernoulli random variable with the probability of success equal to \(p_i\).  

\subsection{Sequential Updating Process} \label{sec2.2}
\begin{figure}
\begin{center}
\includegraphics[width=8.4cm]{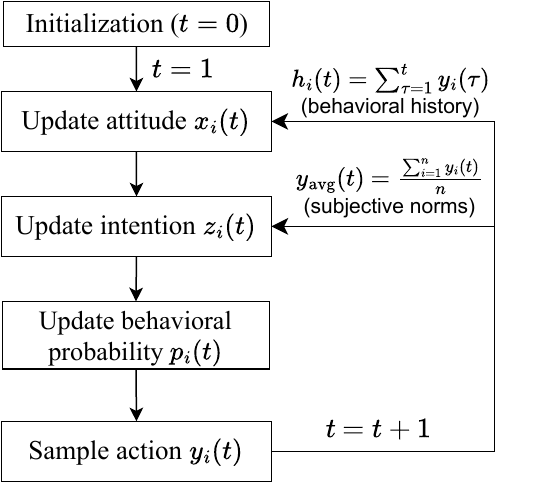}    % The printed column width is 8.4 cm.
\caption{Model process flowchart.} 
\label{fig1}
\end{center}
\end{figure}

The model adopts a sequential updating mechanism, as shown in Fig.~\ref{fig1}, where agents update their state variables in the following order: attitude, intention, behavioral probability, and action. The realized action then feeds back into the next iteration's attitude and intention, thereby forming a dynamic feedback loop between \(x_i, z_i\) and \(y_i\).      
%inspired by Self-Perception Theory (SPT) \citep{SPT} and TPB \citep{TPB}. According to SPT, individuals often infer and adjust their attitudes by observing their own past behaviors \citep{SPT}. Although the theory does not prescribe a quantitative mapping from one's past behavior to attitude, it provides a conceptual foundation for modeling an agent's attitude as a function of its behavioral history. In addition, TPB has established that one's intention is primarily shaped by its attitude and subjective norms, i.e., the person's perceived social pressure to perform or not to perform the behavior \citep{TPB}. As a result, each agent's intention is updated based on its revised attitude and the population-level social norms, represented by the average action across all agents, thereby forming a dynamic feedback loop between \(x_i, z_i\) and \(y_i\).     

\subsubsection{Attitude}
Empirical studies have shown that individuals often adjust their attitudes toward a behavior based on its consequences. For example, a behavior is unlikely to be repeated if the prior consequences of performing it were highly negative \citep{behavior_consequence}. For simplicity, we assume that every time an agent performs a behavior, the resulting consequences shift the agent's attitude \(x_i\) in the same direction, either always increasing or decreasing it. Hence, behaviors can be classified into two types in our model: one with positive consequences that increase \(x_i\) (beneficial type) and the other with negative consequences that decrease \(x_i\) (harmful type). 

Before presenting the attitude updating rule, we first elaborate on its design concept. Psychophysical numbing \citep{psycho_numb_1}, a fundamental principle in the field of psychophysics, captures the phenomenon where people's sensitivity to incremental changes in a physical stimulus declines as the stimulus magnitude increases. For example, experiments have shown that people perceive saving a fixed number of lives as less valuable when the total number of lives at risk become larger \citep{psycho_numb_2}. As discussed earlier, the consequences of performing a behavior can be viewed as a stimulus that shapes the agent's attitude in our model. As the behavior is repeated, the impact of its consequences accumulates, increasing the overall stimulus magnitude. Motivated by the principle of psychophysical numbing, each agent's updated attitude \(x_i(t+1)\) is defined as a function of its latest cumulative behavioral count, denoted \(h_i(t)=\sum_{\tau=0}^t y_i(\tau)\), depending on the behavior type, given by: 
\begin{equation} \label{e1}
x_i(t+1)=
\begin{cases}
1-\dfrac{1-x_i(0)}{1+\lambda \,h_i(t)}, & \text{beneficial type}, \\[8pt]
\dfrac{x_i(0)}{1+\lambda \, h_i(t)}, & \text{harmful type},
\end{cases}
%\;\, t \in \{0,1,2,\dots\},
\end{equation}
where \(x_i(0) \in [0,1]\) is agent \(i\)'s initial attitude, and \(\lambda>0\) is a homogeneous parameter that controls agents' sensitivity of attitude to their cumulative behavioral count. 

\begin{figure}
\begin{center}
\includegraphics[width=8.4cm]{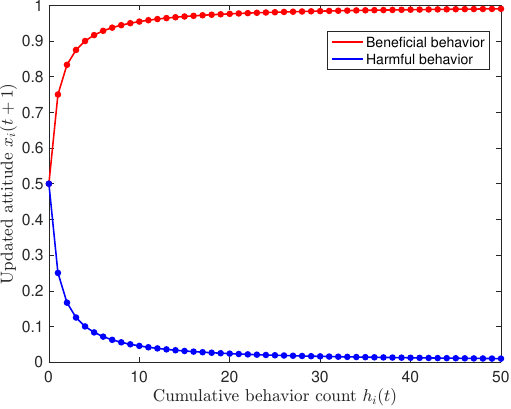}    % The printed column width is 8.4 cm.
\caption{Illustration of attitude as a function of cumulative behavioral count for behaviors with positive and negative consequences, where \(x_i(0)=0.5\) and \(\lambda=1\).} 
\label{fig2}
\end{center}
\end{figure}

As depicted in Fig.~\ref{fig2}, the attitude function is concave for behaviors with positive consequences and convex for those with negative ones. This curvature structure parallels the S-shaped value function introduced in PT \citep{PT}, which is concave for gains and convex for losses. Regardless of behavior type, the marginal change in one's attitude decreases as its cumulative behavioral count grows, consistent with the numbing effect discussed earlier.

\subsubsection{Intention}
As established in TPB \citep{TPB}, the more favorable a person's attitude and subjective norms are toward a behavior, the stronger their intention is to perform it. As mentioned earlier, attitude represents one's internal evaluation of the behavior. In contrast, subjective norms capture one's perceived social pressure imposed by others and consist of two components: injunctive norms and descriptive norms \citep{social_norms}. In particular, individuals form their descriptive norms by observing the behavior of others, which indicates how prevalent the behavior is within the group \citep{TPB_questionnaire}. For simplicity, we use the average action across all agents, i.e., \(y_\mathrm{avg}(t)=\tfrac{1}{n}\sum_{i=1}^n y_i(t) \), to represent the global subjective norms of the group. Since \(y_i(t) \in \{0,1\}\), \(y_\mathrm{avg}(t)\) is equivalent to the proportion of agents who have performed the behavior at time \(t\) (i.e., for whom \(y_i(t)=1\)), thus capturing the prevalence of behavior in the population. Inspired by our earlier work \citep{song_et_al}, each agent's intention is updated as a convex combination of its current attitude and the group's global subjective norms from the previous time step, given by:  
\begin{equation} \label{e2}
z_i(t+1)=\phi \, x_i(t+1)+(1-\phi)\, y_{\mathrm{avg}}(t),
\end{equation}

where \(\phi \in [0,1]\) is a homogeneous parameter that captures the influence weight of agents' attitude on their intention, while \(1-\phi\) reflects their degree of conformity to subjective norms when forming their intention.

\subsubsection{Choice Probability and Action}
Given that \(z_i\) is defined on a normalized scale of \([0,1]\) and each agent makes a binary choice, either to perform or to reject the behavior, the complement \(1-z_i\) can be naturally interpreted as the agent's intention to reject it. To map intention into choice probability, we apply the binary logit choice model \citep{logit}, with \(z_i\) and \(1-z_i\) serving as the utilities of the two alternatives, formulated as:  
\begin{equation} \label{e3}
p_i(t+1)=\frac{\exp\left[\beta z_i(t+1)\right]}{\exp\left[\beta z_i(t+1)\right]+\exp\left[\beta \left(1-z_i(t+1)\right)\right]},
\end{equation}

where \(\beta \geq 0\) is a homogeneous parameter that governs how sensitive agents' choice probabilities are to the utility differences between the two alternatives. It follows directly from~\eqref{e3} that \(p_i\) is a monotonically increasing function of \(z_i\): when \(z_i>1-z_i\), i.e., \(z_i>0.5\), \(p_i\) exceeds \(0.5\); conversely, if \(z_i<0.5\), then \(p_i<0.5\). This property aligns with established socio-psychological principle that the stronger one's intention is, the more likely they will perform the behavior \citep{Ajzen_Fishbean}.   

Although the terminology for \(\beta\) varies across the literature, we interpret it as a measure of agents' rationality in decision-making: when \(\beta=0\), agents randomly choose between the two alternatives with equal probability; as \(\beta \to \infty\), agents select the alternative with higher utility in a deterministic manner. Thus, a smaller value of \(\beta\) implies that agents are less rational, i.e., stronger randomness. 

Finally, each agent's action \(y_i\) is sampled from a Bernoulli distribution with probability of success equal to \(p_i\).   
\begin{equation} \label{e4}
y_i(t+1) \sim \mathrm{Bernoulli}(p_i(t+1))
\end{equation}

This sequential updating process is applied to all agents at every time step. In the next section, we examine the model dynamics through numerical simulations and investigate how the parameters \(\phi\) and \(\beta\) shape social transitions.

\section{Results and Discussion} \label{sec3}
In this section, we present some numerical results of the proposed model. We first specify the initialization schemes for the two behavioral types. Then, we examine the impact of \(\phi\) and \(\beta\) on the collective behavioral transition process for each behavioral type through sensitivity analysis.

\subsection{Initialization} \label{sec3.1}
Findings from behavioral science suggest that individuals are more likely to adopt and sustain behaviors that yield positive consequences, and to abandon those that lead to negative consequences \citep{behavior_consequence}. In light of this distinction, we design different initialization settings corresponding to the two behavioral types. For behavior with positive consequences, most agents are initially set to reject it, allowing us to observe the conditions under which the beneficial behavior is widely adopted. In contrast, for behavior with negative consequences, most agents are initialized to perform it, enabling us to examine when and how they stop the harmful behavior. 

We consider a population of \(n=|V|=300\) agents. Let \(\alpha \in [0.5,1]\) denote the proportion of agents in the majority group, and let \(\mathcal{M} \subseteq V\) be the corresponding index set. Sensitivity analysis has shown that different values of \(\alpha\) only affect the speed of transition without altering its outcome. Thus, we fix \(\alpha\) = 0.9, corresponding to a majority group containing \(|\mathcal{M}|=\alpha n=270\) agents, and examine the impact of \(\phi\) and \(\beta\) on the transition process. 
%In Sec~\ref{sec3.3}, we vary the values of \(\alpha\) to assess how initial agent proportions influence the resulting social transition dynamics. 

\subsubsection{Behavior with Positive Consequences} 
For behavior with positive consequences, we initialize the majority of agents with low attitude and intention values, implying that the behavior is initially not prevalent within the population. Specifically, for each agent \(i \in \mathcal{M}\), their initial attitude and intention are drawn as: \(z_i(0)=x_i(0) \sim U[0,0.4]\), where \(U[a,b]\) denotes the uniform distribution on \([a,b]\). Their initial behavioral probability \(p_i(0)\) and action \(y_i(0)\) are then computed according to~\eqref{e3} and~\eqref{e4}. The remaining minority set of agents, indexed by \(V\setminus\mathcal{M}\), are initialized with higher attitude and intention values. For each agent \(j \in V\setminus\mathcal{M}\), we let \(z_j(0)=x_j(0)\sim U[0.6,0.7]\), with \(p_j(0)\) and \(y_j(0)\) computed from \eqref{e3} and \eqref{e4}.        
\subsubsection{Behavior with Negative Consequences}
For behavior with negative consequences, the initialization setting is contrary to the former case. The majority group is assigned high initial attitude and intention values, indicating that the behavior is widely adopted at the outset. The attitude and intention of each agent are initialized as follows: \(z_i(0)=x_i(0)\sim U[0.6,1], \, \forall \,i \in \mathcal{M}\) and \(z_j(0)=x_j(0)\sim U[0.3, 0.4], \, \forall \, j \in V \setminus \mathcal{M}\). The initial behavioral probability and action are then obtained from \eqref{e3} and \eqref{e4}.   

\subsection{Impact of Attitude Weight \(\phi\) and Irrationality \(\beta\)} \label{sec3.2}
As defined in Sec.~\ref{sec2.2}, the parameter \(\phi \in [0,1]\) specifies the relative influence of agents' attitudes, as opposed to subjective norms, on their intentions. A larger value of \(\phi\) implies that agents place greater weight on their personal evaluation of the behavior and exhibit less conformity to social pressure when forming their intentions. The parameter \(\beta \geq0\), on the other hand, captures the degree of rationality in agents' choices: a smaller value of \(\beta\) introduces stronger randomness into the decision-making process, leading to less rational responses. In addition, the parameter \(\lambda>0\) governs the rate at which attitudes change with behavior occurrence. It has also been observed in sensitivity analysis that \(\lambda\) only affects the transition time. Thus, we fix \(\lambda=1\) throughout the simulations and our main objective is to examine how the two key parameters, attitude influence weight \(\phi\) and decision rationality \(\beta\), shape the collective behavioral transition process.  

In our earlier work \citep{beta_scale}, we inferred the distribution of \(\beta\) from survey data collected among a group of Swedish university students regarding their perception of unlikely events, and obtained an average estimate of \(\beta_{\mathrm{avg}} \approx 8.3\). Guided by this empirical insight, we adopt \(\beta=10\) as the baseline value, representing the standard level of decision rationality among ordinary individuals. Interested readers are referred to \cite{beta_scale} for more details on the empirical distribution of \(\beta\). Thus, we evaluate the outcomes of all combinations of parameter values: \(\phi \in \{0.3,0.7\}\) and \(\beta \in \{5,10\}\). For each simulation scenario, we track the temporal evolution of \(y_\mathrm{avg}(t)\), which reflects the population-level behavior adoption rate, to observe the collective trend.

\begin{figure}
\begin{center}
\includegraphics[width=8.4cm]{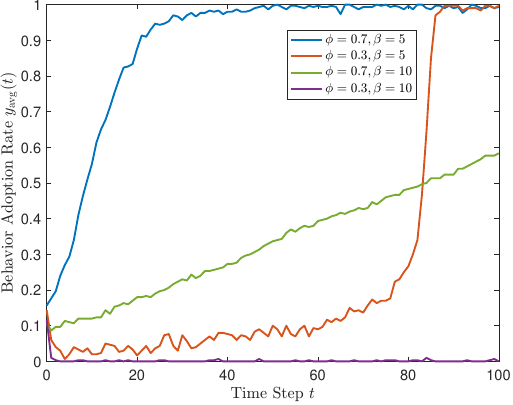}    % The printed column width is 8.4 cm.
\caption{Social dynamics for \(\phi \in \{0.3,0.7\}\) and \(\beta \in \{5,10\}\) (behavior with positive consequences).} 
\label{fig3}
\end{center}
\end{figure}

\subsubsection{Adoption of Beneficial Behavior} We begin by presenting the results for behavior with positive consequences, as shown in Fig.~\ref{fig3}. For the baseline scenario \((\phi,\beta)=(0.7,5)\), the population reaches full adoption by \(t\approx48\) time steps, which is the fastest transition observed among the four scenarios. When either agents' attitude weight \(\phi\) decreases or their rationality level \(\beta\) increases, the social transition slows substantially. Specifically, for \((\phi,\beta)=(0.3,5)\), the complete transition is delayed until \(t \approx 88\). In contrast, for \((\phi, \beta)=(0.7,10)\), the transition proceeds much more slowly and full adoption occurs at \(t\approx300\), which lies outside the plotted range. In the final scenario \((\phi,\beta)=(0.3,10)\), where both the attitude weight \(\phi\) decreases and the level of rationality \(\beta\) increases, no transition occurs and all agents reject the behavior. %Compared with the baseline, these results indicate that both a strong attitude influence and high irrationality level boost the social transition process.

The dynamics observed in Fig.~\ref{fig3} are consistent with the model structure and initial behavioral conditions. Since the majority of agents are initialized to reject the behavior, the initial subjective norm \(y_\mathrm{avg}(0)\) remains low, around \(10\%\) in our setup. In this case, from \eqref{e2}, agents' intention is dominated by their attitude in the early stage. Moreover, since attitude increases with each behavioral occurrence, a large attitude weight \(\phi\) amplifies the resulting positive feedback loop, enabling agents to accumulate intention and expediting the collective transition process. A similar positive feedback mechanism applies to the role of \(\beta\). When the rationality level \(\beta\) is decreased, some agents occasionally perform the behavior despite low intention values. These exploratory actions generate a group of early adopters, which in turn increases the subjective norm of the population and further elevates the intentions of other agents. It seems that a low level of rationality could facilitate adoption of beneficial behavior.      

\begin{figure}
\begin{center}
\includegraphics[width=8.4cm]{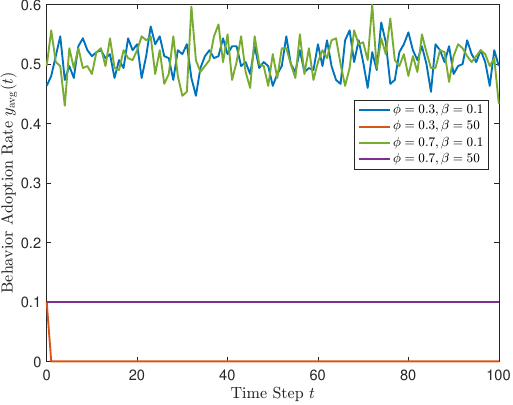}    % The printed column width is 8.4 cm.
\caption{Social dynamics for \(\phi \in \{0.3,0.7\}\) and \(\beta \in \{0.1,50\}\) (behavior with positive consequences).} 
\label{fig4}
\end{center}
\end{figure}

However, the level of rationality cannot be arbitrarily low or high. As illustrated in Fig~\ref{fig4}, when \(\beta=0.1\), corresponding to nearly random decision-making, the behavior adoption rate fluctuates around \(50\%\) irrespective of \(\phi\), and no transition emerges because agents' choices are essentially independent of their intentions. On the contrary, when \(\beta\) is sufficiently large (e.g., \(\beta=50\)), representing fully rational agents (e.g., robots), the transition also fails to occur: the adoption rate either remains at the initial \(10\%\) or declines to zero, depending on the value of \(\phi\). The combined results of Fig.~\ref{fig3} and~\ref{fig4} indicate that although attitude contributes to the transition dynamics, the level of rationality plays a more decisive role. In particular, a moderate level of randomness in decision-making is essential for initiating the adoption of beneficial behaviors, underscoring the importance of ``bounded rationality'' \citep{bounded_rationality} in inducing positive social change. 

The observed results can also be interpreted through the lens of behavioral economics, particularly the notion of ``behavioral nudges'' introduced in the nudge theory \citep{nudge_theory}. Nudge theory suggests that individuals can be guided toward a particular choice by making it more accessible while still preserving people's freedom of choice. For example, people sometimes choose junk food simply because it is convenient and cheap, while healthier options are more expensive and harder to find. In this case, a nudge could be placing healthy food in a more visible location in the supermarket. In general, a nudge serves as a subtle trigger that encourages individuals to overcome their status quo bias and adopt a new, potentially more beneficial behavior. This mechanism reflects the fact that people often rely on quick, impulsive judgments to select the easiest available option, even when doing so is not in their best interest. Therefore, our results are consistent with the behavioral-economic perspective, showing that agents' bounded rationality can lead them to adopt more beneficial behaviors when the environment effectively nudges them toward these options.  

\subsubsection{Rejection of Harmful Behavior}
For behavior with negative consequences, the impact of \(\phi\) and \(\beta\) on agents' collective behavioral transition is similar to the former case, with a key distinction that bounded rationality can either support or hinder the transition depending on the group's initial conditions. 

\begin{figure}
\begin{center}
\includegraphics[width=8.4cm]{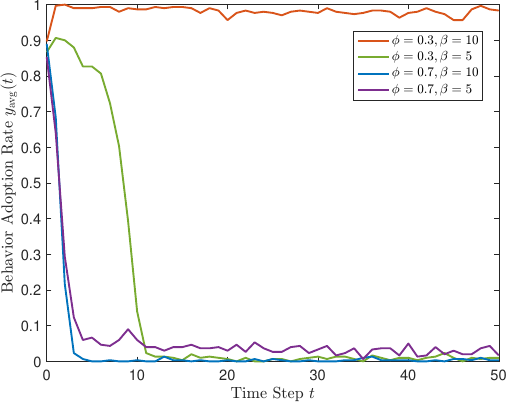}    % The printed column width is 8.4 cm.
\caption{Social dynamics for \(\phi \in \{0.3,0.7\}\) and \(\beta \in \{5,10\}\) (behavior with negative consequences).} 
\label{fig5}
\end{center}
\end{figure}

As shown in Fig.~\ref{fig5}, when \((\phi,\beta)=(0.3,10)\), most agents do not reject the behavior. In this case, increasing \(\phi\) from \(0.3\) to \(0.7\) enables the transition process. Similarly, by comparing the scenarios \((\phi,\beta)=(0.7,5)\) with \((\phi,\beta)=(0.3,5)\), it can be observed that increasing \(\phi\) leads to a faster transition. Thus, a large attitude weight \(\phi\) promotes rejection of the harmful behavior mainly because the initial subjective norm is extremely large, with \(y_\mathrm{avg}(0)\approx 0.9\). From~\eqref{e2}, a larger \(\phi\) implies that agents place less weight on the strong normative pressure, which lowers their intention values and supports the transition away from the behavior. The observed results correspond to the conformity effects demonstrated in Asch's experiment \citep{Asch}, where each participant was surrounded by a unanimous group deliberately making incorrect decisions that contradicted the truth. Most participants conformed to the normative pressure, while only a few maintained their independent evaluation and selected the correct option. These empirical findings highlight the situations in which social norms override personal judgment, which can be detrimental when the prevailing norm is misleading. In such cases, it is necessary to enhance the influence of personal attitudes, captured by increasing \(\phi\) in our model, to help individuals resist the harmful behavior. 

The level of rationality \(\beta\), on the other hand, exhibits a fundamentally two-sided effect. When \((\phi,\beta)=(0.3,10)\), reducing \(\beta\) from \(10\) to \(5\) facilitates the transition process. In contrast, when \((\phi,\beta)=(0.7,10)\), the same reduction of \(\beta\) impedes both the speed and completeness of the transition, preventing the population from fully rejecting the behavior. The seemingly contradictory results stem from the different initial conditions induced by \(\phi\). As discussed earlier, when \(\phi\) is small, agents' initial intentions remain high due to the strong subjective norms. In this case, introducing greater randomness into decision-making by lowering \(\beta\) enables some agents to deviate from the adverse behavioral pattern, which further reduces the subjective norm. When \(\phi\) is large, however, most agents already tend to reject the behavior, and extra randomness increases the likelihood that some agents revert to the harmful behavior, thereby impeding the overall transition. In summary, bounded rationality plays a dual role: it may accelerate behavioral change or disrupt it, depending on the system’s initial conditions.     

%\subsection{Impact of Initial Agent Proportion \(\alpha\)} \label{sec3.3}
%Finally, we test the impact of initial proportion of majority agents \(\alpha\) on the transition process. For the beneficial behavior and fixed parameter setting \((\phi,\beta)=(0.3,5)\), we vary \(\alpha \in \{0.9,0.8,0.7,0.6\}\) and observe the results. As shown in Fig.~\ref{fig6}, the initial proportion only affects the speed of transition: larger values of \(\alpha\) delay the process but do not alter the eventual outcome. In other words, the system consistently reaches full adoption, but higher initial proportion of the majority increases the transition time.

%\begin{figure}
%\begin{center}
%\includegraphics[width=8.4cm]{ifacconf_latex/result_4.pdf}    % The printed column width is 8.4 cm.
%\caption{Social dynamics for \(\alpha \in \{0.9,0.8,0.7,0.6\}\) when \(\phi=0.3\) and \(\beta=5\) (behavior with positive consequences).} 
%\label{fig6}
%\end{center}
%\end{figure}

%These results indicate that \(\phi\) and \(\beta\) are the primary drivers of collective behavioral change in our model, while \(\alpha\) plays a secondary role by modulating transition speed.

\section{Conclusions and Future Work} \label{sec4}
In this study, we proposed a dynamic agent-based model conceptually rooted in the TPB and combined with a feedback mechanism from behavior to attitude. Specifically, we defined two types of behavior based on their feedback effects and observed two social transition processes, namely the adoption of beneficial behaviors and rejection of harmful ones. Simulation results indicate that the collective behavioral transitions can be controlled by adjusting two key parameters: \(\phi\) and \(\beta\), which characterize agents' attitude influence weight on intention and their decision rationality, offering insights into the mechanisms underlying collective human social behavior.

Two main challenges remain for extending our model into practical applications. First, a realistic measure is needed to estimate the two behavioral parameters at both the individual and population level. The attitude influence weight \(\phi\) could be inferred through TPB-based surveys that evaluate the extent to which individuals commit to their personal attitudes when making decisions. Alternatively, the decision rationality \(\beta\) could be estimated from prospect choice experiments, as exemplified by the rationality index developed by \cite{demartino}, or from surveys that assess individuals' perception of unlikely events, as shown in \cite{beta_scale}. Second, effective strategies are needed to induce parameter changes at the population level. For instance, behavioral interventions that strengthen personal beliefs may increase \(\phi\), while the framing effect \citep{demartino} can be used to adjust \(\beta\). Our future work will focus on bridging these gaps and applying the model to real-world case studies.        

\bibliography{main}             % bib file to produce the bibliography

@article{TPB,
	author={I. Ajzen},
	title={The theory of planned behavior},
	journal={Organ. Behav. Hum. Decis. Process.},
	year={1991},
	volume={50(2)},
	pages={179--211},
}

@article{TPB_application,
	author={Hongyun Si and Jiangang Shi and Daizhong Tang and Shiping Wen and Wei Miao and Kaifeng Duan},
	title={Application of the theory of planned behavior in environmental science: A comprehensive bibliometric analysis},
	journal={Int. J. Environ. Res. Public Health},
	year={2019},
	volume={16(15)},
	pages={2788},
}

@article{PT,
	author={D. Kahneman and A. Tversky},
	title={Prospect theory: An analysis of decision under risk},
	journal={Econometrica},
	year={1979},
	volume={47(2)},
	pages={263--291},
}

@techreport{TPB_questionnaire,
    author = {I. Ajzen},
    title = {Constructing a theory of planned behavior questionnaire},
    institution = {UMass Amherst},
    year = {2006}
}

@article{social_norms,
	author={Franziska Heinicke and Christian König-Kersting and Robert Schmidt},
	title={Injunctive vs. descriptive social norms and reference group dependence},
	journal={J. Econ. Behav. Organ.},
	year={2022},
	volume={195},
	pages={199--218},
}

@article{song_et_al,
  author={Song, Chen and Cvetkovic, Vladimir and Su, Rong}, 
  title={Why Do Opinions and Actions Diverge? A Dynamic Framework to Explore the Impact of Subjective Norms},
  journal={IEEE Transactions on Computational Social Systems},
  year={2025},
  pages={1-12},
  doi={10.1109/TCSS.2025.3598697}
}

@incollection{logit,
    author = {Daniel McFadden},
	title = {Conditional logit analysis of qualitative choice behavior},
    pages = {105--142},
    editor = {Paul Zarembka},
    publisher = {Academic Press},
    year = {1974},
	booktitle = {Frontiers in {Econometrics}},
    address = {New York, NY, USA},
}

@article{Ajzen_Fishbean,
	author = {Martin Fishbein and Icek Ajzen},
    title = {Belief, attitude, intention, and behavior: An introduction to theory and research},
	journal = {Philos. Rhetor.},
    year = {1977},
    volume = {10(2)},
	pages = {130--132},
}

@article{psycho_numb_1,
	author={D. Fetherstonhaugh and P. Slovic and S. M. Johnson and J. Friedrich},
	title={Insensitivity to the value of human life: A study of psychological numbing},
	journal={J. Risk Uncertain.},
	year={1997},
	volume={14(3)},
	pages={283--300},
}

@article{psycho_numb_2,
	author={J. Friedrich and P. Barnes and K. Chapin and I. Dawson and V. Garst and D. Kerr},
	title={Psychophysical numbing: When lives are valued less as the lives at risk increase},
	journal={J. Consum. Psychol.},
	year={1999},
	volume={8(3)},
	pages={277--299},
}

@book{nudge_theory,
	author={R. H. Thaler and C. R. Sunstein},
	title={Nudge: Improving Decisions About Health, Wealth, and Happiness},
	publisher={Yale University Press},
	year={2008},
	address={New Haven, CT, USA},
}

@article{beta_scale,
	author={A. Fontan and V. Cvetkovic and P. Herman and J. Sundh and K. H. Johansson},
	title={Exploring rationality of prospect choices among decision-makers in a population},
	journal={IFAC-PapersOnLine},
	year={2024},
	volume={58(30)},
	pages={133--138},
}

@article{bounded_rationality,
author = {H. A. Simon},
title = {A Behavioral model of rational choice},
journal = {Q. J. Econ.},
year = {1955},
volume = {69(1)},
pages = {99--118}
}

@incollection{Asch,
	author={S. E. Asch},
	title={Effects of group pressure upon the modification and distortion of judgments},
	booktitle={Groups, Leadership and Men: Research in Human Relations},
    editor ={H. Guetzkow},
	publisher={Carnegie Press},
	year={1951},
	pages={177--190},
	address={Pittsburg, PA, USA},
}

@book{behavior_consequence,
	author={B. F. Skinner},
	title={Science and Human Behavior},
	publisher={Macmillan},
	year={1953},
	address={New York, NY, USA},
}

@book{cognitive_dissonace,
	author={Joel Cooper},
	title={Cognitive Dissonance: Fifty Years of a Classic Theory},
	publisher={SAGE Publications Ltd},
	year={2007},
	address={London, UK},
}

@article{demartino,
	author={Benedetto {De Martino} and Dharshan Kumaran and Ben Seymour and Raymond J. Dolan},
	title={Frames, biases, and rational decision-Making in the human brain},
	journal={Science},
	year={2006},
	volume={313(5787)},
	pages={684--687},
}
                                                     % with bibtex (preferred)
\end{document}